\documentclass{aastex}
\usepackage{spr-astr-addons}
\usepackage{url}

\usepackage{subfigure}

\RequirePackage{color}

\begin{document}

\title{Experimental Studies of Magnetically Driven Plasma Jets}
\slugcomment{Not to appear in Nonlearned J., 45.}
\shorttitle{Experimental Studies of Magnetically Driven Plasma Jets}
\shortauthors{F. Suzuki-Vidal et al.}

\author{F. Suzuki-Vidal} \and \author{S. V. Lebedev} \and \author{S. N. Bland} \and \author{G. N. Hall} \and \author{G. Swadling} \and \author{A. J. Harvey-Thompson} \and  \and \author{G. Burdiak} \author{P. de Grouchy} \author{J. P. Chittenden} \and \author{A. Marocchino} \and \author{M. Bocchi}
\affil{The Blackett Laboratory, Imperial College London, London, UK.}
\and
\author{A. Ciardi}
\affil{LERMA, Universit{\'e} Pierre et Marie Curie, Observatoire de Paris and {\'E}cole Normale Sup{\'e}rieure. UMR 8112 CNRS.}
\and
\author{A. Frank}
\affil{Department of Physics and Astronomy, University of Rochester, Rochester, NY, USA.}
\author{S. C. Bott}
\affil{Center for Energy Research, University of California, San Diego, CA, USA.}

\email{f.suzuki@imperial.ac.uk}

\begin{abstract}
We present experimental results on the formation of supersonic, radiatively cooled jets driven by pressure due to the toroidal magnetic field generated by the 1.5 MA, 250 ns current from the MAGPIE generator. The morphology of the jet produced in the experiments is relevant to astrophysical jet scenarios in which a jet on the axis of a magnetic cavity is collimated by a toroidal magnetic field as it expands into the ambient medium. The jets in the experiments have similar Mach number, plasma beta and cooling parameter to those in protostellar jets. Additionally the Reynolds, magnetic Reynolds and Peclet numbers are much larger than unity, allowing the experiments to be scaled to astrophysical flows. The experimental configuration allows for the generation of episodic magnetic cavities, suggesting that periodic fluctuations near the source may be responsible for some of the variability observed in astrophysical jets. Preliminary measurements of kinetic, magnetic and Poynting energy of the jets in our experiments are presented and discussed, together with estimates of their temperature and trapped toroidal magnetic field.
\end{abstract}



\section{Introduction}

Collimated outflows (jets) are associated with widely diverse astrophysical environments but exhibit many common features which are independent of the central source (\cite{livio02nature}). In general, it is believed that the ejection of jets relies on the conversion of gravitational energy into Poynting flux which powers the outflows. The standard magneto-hydrodynamic (MHD) models of jet formation rely on differential rotation along a large scale poloidal magnetic field $B_{P}$ to generate a toroidal magnetic field component $B_{\phi}$ which accelerates and collimates a disk-wind (\cite{blandford82}). Our experiments are designed to model the acceleration and collimation of astrophysical jets taking place under the condition $|B_{\phi}|$$\gg$$|B_{P}|$.

Astrophysical jets and outflows are described to a first approximation by ideal MHD, which is a valid approximation when the dimensionless Reynolds ($Re$), magnetic Reynolds ($Re_M$), and Peclet ($Pe$) numbers are much larger than unity (\cite{ryutov99, ryutov00}). In this regime the transport of momentum, magnetic fields, and thermal energy, respectively, occur predominantly through advection with the flow. 

In this paper we wish to extend the study of supersonic (Mach number$>$1), radiatively cooled, magnetically driven (plasma beta, the ratio of the thermal to magnetic pressure, $\beta$$\lesssim$1) plasma jets from laboratory experiments which are relevant to the launching mechanism in astrophysical jet models (\cite{suzukividal09, ciardi09, lebedev05mnras}). We present new experimental results from radial foils which include the energy balance inside a magnetic cavity, temperature of the jet and trapped magnetic field inside the outflows.

\section{Experimental setup}

\begin{figure}[t]
\begin{center}
  {\includegraphics[width=7.5cm]{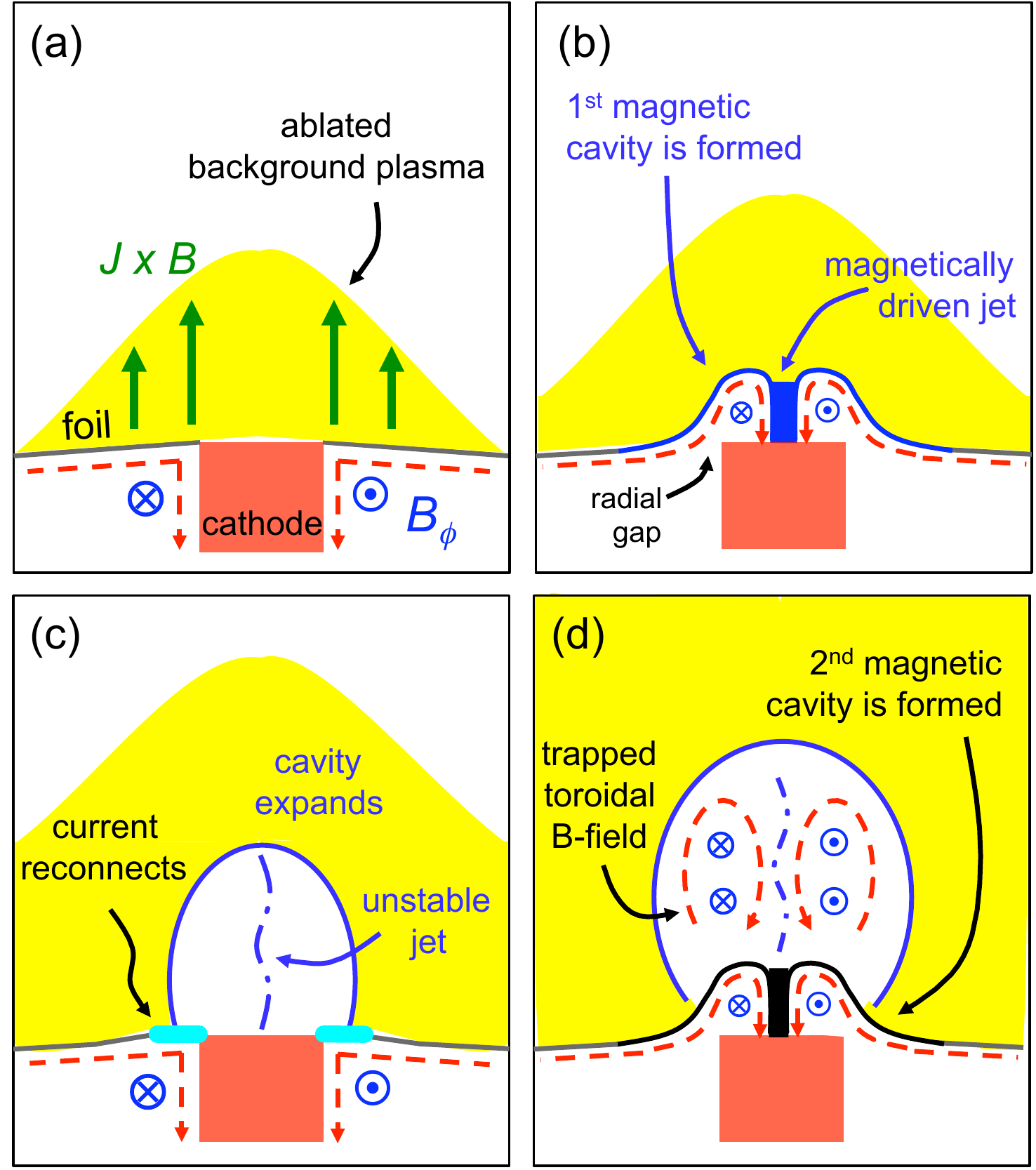}}
  \end{center}
  \caption{Schematic evolution of the foil showing the mechanism of episodic magnetic cavity formation triggered by current reconnection at the base of the cathode. The figure shows the current path (red-dashed arrows), toroidal magnetic field (blue arrows, pointing perpendicular to the page) and the resultant $J\times B$ force (green arrows).}
\label{fig:foil_setup}
\end{figure}

Our experiments use a radial aluminium foil with a diameter of $\sim$55 mm and a typical thickness of 6.5 $\mu$m, held in place between two concentric electrodes. The central cathode is a hollow stainless steel cylinder with a diameter of $\phi_c$=3.1 mm in contact with the foil at its centre. The foil is the load of the 1 MA, 250 ns (zero to maximum) current pulse from the MAGPIE generator (\cite{mitchell96}). Fig.~\ref{fig:foil_setup}a shows a schematic side-on view of the radial foil. The central cathode together with the current path, toroidal magnetic field $B_{\phi}$ and resultant Lorentz $J\times B$ force, which is normal to the surface of the foil, are indicated in the figure. This configuration results in a current path along the central cathode, radially along the foil and along a return-current structure (not shown in Fig.~\ref{fig:foil_setup}). The toroidal magnetic field for peak current at the radius of the cathode is $B_{\phi}(r,t)$=$\mu_0 I(t)/2 \pi r$$\sim$100 T ($=1$ MG). 

The dynamics of the jet formation were studied with several plasma diagnostics: optical laser probing ($\lambda$=532 nm, $\sim$0.4 ns exposure) was used for shadowgraphy, schlieren and interferometry; XUV self-emission ($h\nu$$>$30 eV) from the plasma was recorded using multi-frame, time-resolved ($\sim$3 ns gate) pinhole imaging providing up to 8 frames per experiment; Photo-Conducting Detectors (PCDs) were used to measure the X-ray emission; magnetic ``B-dot'' probes measured any trapped magnetic field inside the outflows; an inductive probe connected to the cathode measured the voltage, and thus Poynting flux driving the outflows; an X-ray spectrometer with a spherically bent mica crystal allowed measuring the temperature of the jet. 

\begin{figure*}[t]
\begin{center}
    {\includegraphics[width=15cm]{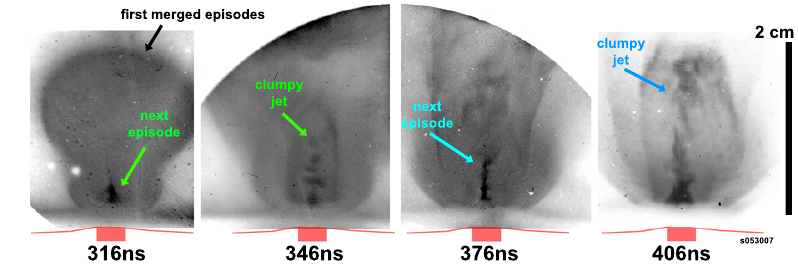}}
\end{center}	
\caption{Late-time evolution of the episodic magnetic cavities from 316 ns onwards. The cavities emerge inside the remnants left by the earlier episodes.}
  \label{fig:s053007long}
\end{figure*}

\section{Evolution of a magnetically driven jet}

The time evolution of a radial foil, leading to the formation of a typical experimental magnetically driven jet/outflow, are summarized schematically in Figs.~\ref{fig:foil_setup}a-d. Two outflow components are present: a magnetic bubble (or cavity) accelerated by gradients of the magnetic pressure and surrounded by a shell of swept-up ambient material, and a magnetically confined jet in the interior of the bubble. The rate of expansion of the magnetic cavity itself depends on the density of an external medium provided by the plasma ablated from the foil at early times, prior to the formation of the cavity and the jet. Although the dynamics of the first magnetic bubble and jet are similar to those observed in radial wire array jet experiments (\cite{lebedev05mnras, ciardi07pop}), use of a foil produces an episodic jet activity. Our major departure from previous experiments is the foil mass now increases with radius, where it remained constant for radial wire arrays. One consequence of this additional mass is that the initial radial gap (Fig.~\ref{fig:foil_setup}b) produced by the full ablation of foil material at the cathode radius is smaller and can be more easily refilled by the plasma expanding from the foil and/or the cathode. The gap closure allows the current to flow once again across the base of the magnetic cavity, thus re-establishing the initial configuration. When the magnetic pressure is large enough to break through this newly deposited mass, a new jet/bubble ejection cycle begins. The temporal evolution of the episodic jets and cavities at later times (after $\sim$3 ejections) is presented in Fig.~\ref{fig:s053007long}, where these images were obtained during the same experiment. A succession of multiple cavities and embedded jets is seen. The resulting flow is heterogeneous and \emph{clumpy}, and it is injected into a long-lasting and well collimated channel made out of nested cavities. It is worth remarking that the bow-shaped envelope is driven by the magnetic field and not hydrodynamically by the jet.

\section{Latest experimental results}

\subsection{Kinetic energy}

\begin{figure}[h!]
\begin{center}
	\subfigure{\includegraphics[width=5.5cm]{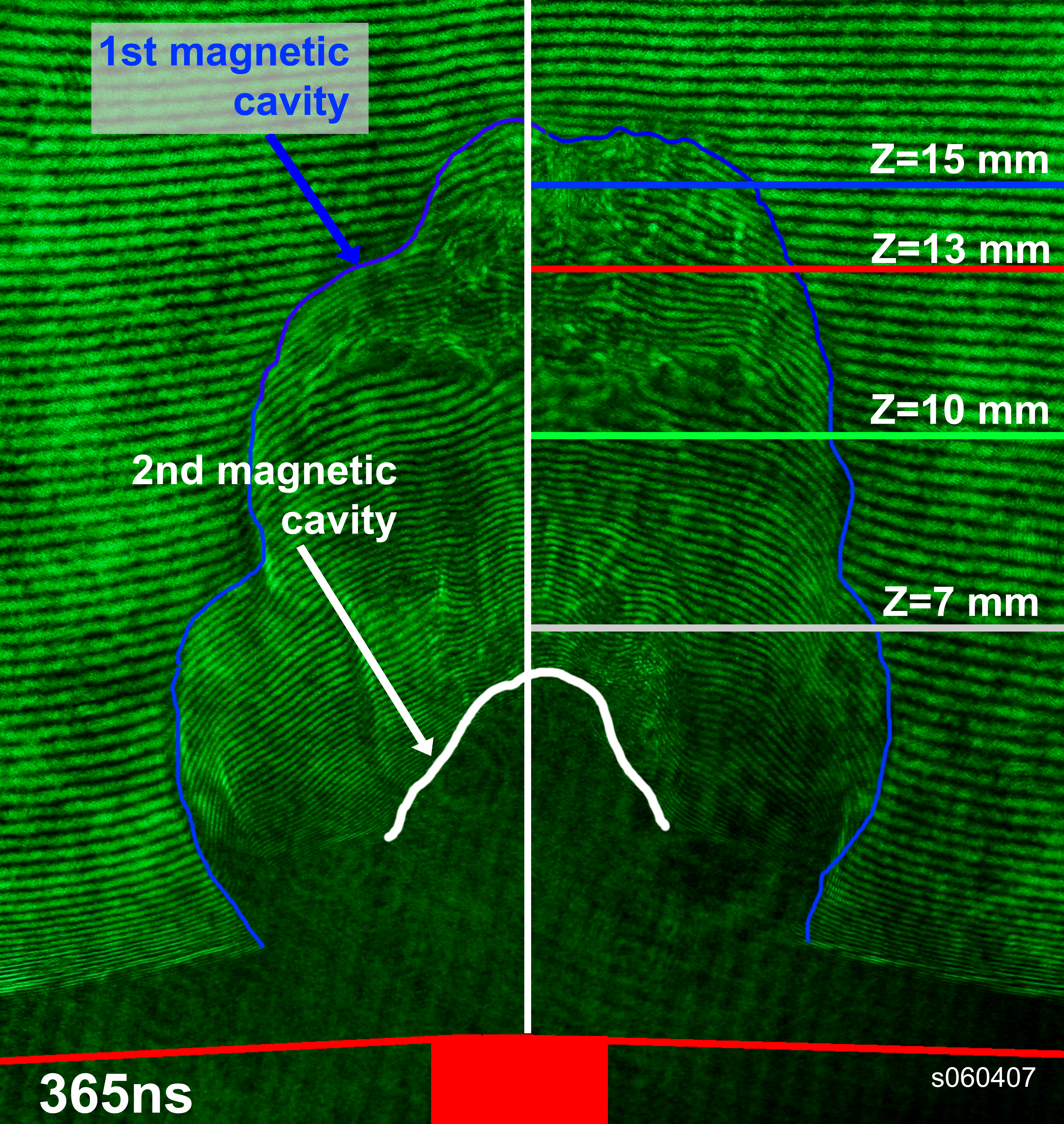}}
\end{center}	
\caption{Side-on laser interferogram at 365 ns, with the position of the foil and the cathode shown schematically. Two magnetic cavities can be seen with their contours highlighted. Also shown are the radial profiles used to obtain the mass inside the cavity.}
  \label{fig:kinetic}
\end{figure}

The amount of mass inside a magnetic cavity was estimated by measuring the distribution of electron density of the plasma using laser interferometry. An example of interferometry data is shown in Fig.~\ref{fig:kinetic}, obtained at 365 ns and illustrating the formation of two subsequent magnetic cavities. The electron density integrated along the line of sight was obtained as a function of radius at different heights from the foil ($n_el(r,z)$, shown schematically on the image). The mass inside the magnetic cavity was estimated by integrating the radial profiles of $n_el(r,z)$, having assumed a value for the ionization of the plasma of $Z$=5. The plasma ionization was estimated from measurements of the electron temperature of the jet in Sec.~\ref{temp}. The radial contours in Fig.~\ref{fig:kinetic} show typical values of $n_el$$\sim$10$^{18-19}$ cm$^{-2}$, which resulted in a total mass of aluminium inside the cavity of $m$$\sim$10$^{-8}$ kg.

To a first approximation the distribution of the velocities in the plasma inside the magnetic cavity can be assumed to increase linearly with height. This assumption is based on the motion of different features along the jet, as seen also in the jets from single-episode magnetic cavities in radial wire arrays (\cite{suzukividal10_ieee_tps}). The axial tip velocity of the first cavity was measured from time-resolved XUV emission images, resulting in $V_{Ztip}$$\approx$128 km/s. The kinetic energy inside the cavity as a function of height can then be estimated as $E_K(z)$$\propto$$\frac{1}{2} m(z)~V_{Ztip}^2$, resulting in values of $E_K$$\sim$50-160 J.  

\subsection{Magnetic energy and injected Poynting flux}

\begin{figure}[t]
\begin{center}
\includegraphics[width=8cm]{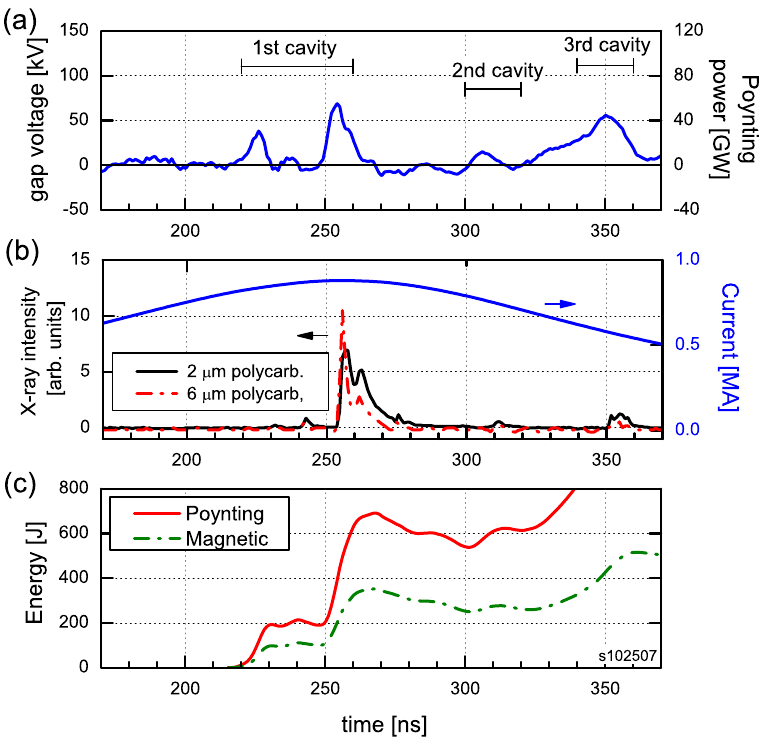}
\end{center}
\caption{(a) Voltage from an inductive probe and Poynting power on the right-hand side scale, (b) X-ray emission and load current. The signals from both diagnostics are correlated to the formation of subsequent magnetic cavities. (c) Magnetic and Poynting energy obtaind from (a).}
\label{fig:energy}
\end{figure}

The amount of energy delivered from the generator to the magnetic cavity and jet was estimated using an inductive probe connected to the cathode. This diagnostic measures the voltage drop produced by the current path along the cathode, jet, and walls of the magnetic cavity. The probe voltage $V_{ind}$ is proportional to the time derivative of the magnetic flux produced by the current ($V_{ind}$=$d(LI)/dt$), where $L$ is the inductance associated with the current path carrying total current $I$. An example of these measurements is shown in Fig.~\ref{fig:energy}a as a series of rapid changes in voltage (seen at $\sim$220, 250 and after 300 ns). These voltage spikes are coincident in time with X-ray emission bursts shown in Fig.~\ref{fig:energy}b, which are in turn correlated to the formation of each jet/cavity. The voltage from the inductive probe reflects the additional (time-dependent) inductance of the jet and the magnetic bubble. Assuming that all the current flows inside the cavity, it is possible to calculate how the inductance changes in time by integrating the results from Fig.~\ref{fig:energy}a. We estimate that the inductance of the cavity is $L$$\sim$1 nH, which is consistent with the geometrical inductance from the size of the cavity from imaging diagnostics. Fig.~\ref{fig:energy}c shows estimates for the magnetic energy inside the cavity, resulting in $E_M$=$LI^2/2$$\sim$500 J. It is also possible to estimate the total energy delivered to the cavity by the Poynting flux, by integrating the Poynting power $V_{ind}I$ (shown also in Fig.~\ref{fig:energy}a), resulting in $E_P$$\sim$600-800 J. The difference between $E_M$ and $E_P$ will be shared between kinetic energy of the outflow, radiation and internal energy of the plasma. Further details on these measurements will be presented in future publications.

\subsection{Temperature of the jet}
\label{temp}

Obtaining the jet temperature is important for estimating the magnetic Reynolds number. The temperature of the plasma in the magnetic cavity was measured using time-integrated hard X-ray spectroscopy (photon energy of $h\nu$$>$1 keV), which provided information about the hot, compressed part of the outflow. The spectrum was measured using a spherically bent mica crystal (\cite{hall06pop}), which provided spatial resolution along the length of the jet. The electron temperature of the jet $T_e$ was estimated by considering the ratio of the combined intensity of Helium-like aluminium lines to Hydrogen-like aluminium. The dependance of these line ratios on temperature has been modeled by \cite{apruzese97} for a plasma with parameters close to our experiment. In this way we determine the temperature of the jet as a function of height, resulting in values of $T_e$$\sim$170-600 eV. 

From these temperature estimates it is possible to infer characteristic values of the magnetic Reynolds number $Re_M$ of the plasma, given by $Re_M$=$LV/D_M$, where $L$ and $V$ are the characteristic length and velocity of the plasma, and $D_M$ is the magnetic diffusivity. $D_M$ can be expressed as a function of the temperature $T$ and ionization state $Z$ of the plasma (\cite{ryutov00}). For our experimental conditions we can estimate $Re_M$ for both the central jet and magnetic cavity if we assume a typical flow velocity of $V$$\sim$100 km/s. For the central jet we take the measured temperatures of $T$=170-600 eV, jet diameter of $L$$\sim$1 mm, and $Z$$\sim$5-10 (estimated from the measured temperatures assuming ionization balance), resulting in a range of $Re_M$$\approx$300-1000. These estimates show that the magnetic Reynold numbers that characterize the plasma jet and magnetic cavity are significantly larger than unity and thus in the regime relevant for scaled representations of astrophysical outflows. 

\subsection{Trapped magnetic field}

\begin{figure}[h!]
\begin{center}
	{\includegraphics[height=4.8cm]{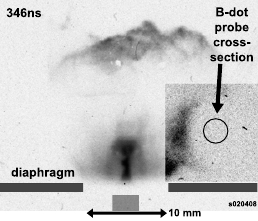}}
 \end{center}
  \caption{XUV emission at 346 ns, showing schematically the setup used to measure any trapped toroidal magnetic field inside the expanding cavities. The image contrast near the position of the magnetic probe has been enhanced in order to show its interaction with the magnetic cavity.}
  \label{fig:bdotsetup}
\end{figure}

The above estimates indicate that the magnetic Reynolds number is much greater than unity and thus some magnetic flux should remain trapped inside the outflows. In particular we can expect conservation of magnetic flux accumulated in the first cavity by the time the second cavity starts to form. The presence of toroidal magnetic field inside the expanding magnetic cavities was measured using a magnetic probe. The probe had a detection area with a diameter of $\sim$3 mm, placed at a radius of $\sim$13 mm from the axis. To exclude capacitive coupling between the probe and the cathode a stainless steel diaphragm was installed above the foil. The magnetically driven jets formed through a 10 mm diameter aperture in the diaphragm, without affecting the dynamics of the episodic jet generation. An XUV emission image at 346 ns in Fig.~\ref{fig:bdotsetup} shows two magnetic cavities expanding above the metallic diaphragm. The position of the magnetic probe together with the cathode, the diaphragm and the circular aperture on the axis are shown schematically in the figure. The voltage measured by the magnetic probe is zero until $\sim$350 ns, which is consistent with the time when the magnetic cavity reaches the probe. The signal from the magnetic probe corresponds to $B_{\phi~trap}$$\sim$0.3 T. This estimate can be compared with the expected magnetic field assuming that the initial toroidal magnetic flux in the first cavity is conserved as the cavity expands and the central jet is pinched, which is of the order of $B_{\phi}$$\sim$1.7 T. The resultant magnetic field is of the right magnitude, though an accurate estimate of the expected field is not possible due to the uncertainty of the current path through the plasma at the time when the cavity reaches the magnetic probe.

\section{Effect of instabilities on the dynamics of episodic magnetic cavities}

\begin{figure}[t]
\begin{center}
	     {{\includegraphics[width=8.1cm]{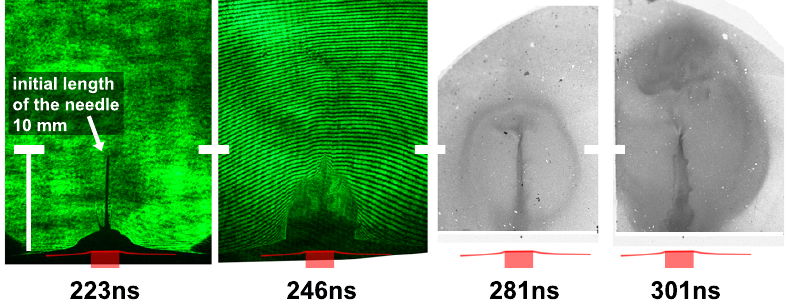}}}
     \end{center}
\caption{Laser probing and XUV emission images (obtained from the same experiment) of a radial foil with a needle on the axis, showing the formation of a single magnetic cavity.}
  \label{fig:needle_xuv_laser}
\end{figure}

It is seen from Fig.~\ref{fig:s053007long} that the jet on the axis of the magnetic cavity is unstable, similar to the situation discussed in our previous experiments with the generation of a single episode magnetically driven jet (\cite{lebedev05mnras, ciardi07pop}). The development of instabilities and subsequent disruption of the jet could contribute to the increase of voltage at the base of the cavity and affect the process of reconnection of current through the gap between the cathode and the remnants of the foil which is responsible for the formation of new jet episodes. To investigate the possible effect of instabilities on the periodicity of the generation of magnetic cavities, a 200 $\mu$m diameter, 10 mm long stainless-steel needle was placed on the axis of the system (above the foil) thus providing a fixed current path on the axis. 

Results from side-on laser probing and XUV emission images obtained from the same experiment are shown in Fig.~\ref{fig:needle_xuv_laser}. The experimental data show that periodic behavior is now suppressed - only a single magnetic cavity is formed in this case. Its dynamics are qualitatively similar to the case without the needle on the axis. The plasma on axis is localized along the length of the needle at early times, becoming unstable after the cavity expands further along its tip. The measured axial and radial velocities of the cavity resulted in $V_Z$$\sim$340 km/s and $V_R$$\sim$124 km/s respectively, which are $\sim$40-50 $\%$ larger than without  stabilization of the current with the needle. The increase in the expansion velocity of the magnetic cavity suggests that a larger fraction of the current is localized inside the single magnetic cavity, thereby increasing the magnetic pressure inside it and hence the expansion velocity.

\section{Summary}

We have presented new experimental data from radial foil experiments which complement previous results and enhance our understanding on the dynamics of episodic magnetically driven jets. Preliminary measurements of the kinetic and magnetic energy inside the magnetic cavities are in good agreement with the total electromagnetic energy injected as Poynting flux, measured with an inductive probe connected at the base of the cavity. The temperature of the jet/outflow, determined by time-integrated spectroscopy, resulted in $T$$\sim$170-600 eV, providing a first estimate of this parameter. These high temperatures are consistent with measurements of toroidal magnetic field inside the cavity, indicating that magnetic flux is conserved as the cavity expands and that the plasma has a magnetic Reynolds number larger than unity. Our results indicate values of the order of $Re_M$$\sim$300-1000, consistent with previous results from computer simulations (\cite{ciardi09}) thus validating the scaling of these experiments to astrophysical jets. In addition we find that episodic jet formation is halted by a metal needle on the axis of the foil. We observed the formation of only a single magnetic cavity, as in our previous experiments with radial wire arrays. This supports the hypothesis that instabilities in the jet may disrupt the current path inside the cavity, encouraging the reconnection of current at the base. Future experiments will look into the detailed dynamics of this new setup.  

\acknowledgments

This work was supported by the EPSRC Grant No. EP/G001324/1 and by the NNSA under DOE Cooperative Agreements No. DE-F03-02NA00057 and No. DE-SC-0001063.

\end{document}